\documentclass[preprint,showpacs,preprintnumbers,superscriptaddress,amsmath,amssymb]{revtex4-2}
\usepackage{amsmath}
\usepackage{natbib}
\usepackage{amssymb}
\usepackage{color,soul}
\usepackage{graphicx}
\usepackage{dcolumn} 
\usepackage{bm} 
\usepackage{hyperref}
\usepackage{longtable}
\usepackage{gensymb}
\usepackage{siunitx}
\usepackage[version=4]{mhchem}
\usepackage{tabularx}
\usepackage[normalem]{ulem}
\begin{document}
\texttt{}

\title{Universal Structural Influence on the 2D Electron Gas at SrTiO$_3$ Surfaces}

\author{Eduardo~B.~Guedes}
\affiliation{Photon Science Division, Paul Scherrer Institut, CH-5232 Villigen, Switzerland}
\affiliation{Institut de Physique, \'{E}cole Polytechnique F\'{e}d\'{e}rale de Lausanne, CH-1015 Lausanne, Switzerland}
\author{Stefan~Muff}
\affiliation{Photon Science Division, Paul Scherrer Institut, CH-5232 Villigen, Switzerland}
\affiliation{Institut de Physique, \'{E}cole Polytechnique F\'{e}d\'{e}rale de Lausanne, CH-1015 Lausanne, Switzerland}
\author{W. H. Brito}
\affiliation{Departamento de F\'{i}sica, Universidade Federal de Minas Gerais, C.P. 702, 30123-970 Belo Horizonte, Minas Gerais, Brazil}
\author{Marco~Caputo}
\affiliation{Photon Science Division, Paul Scherrer Institut, CH-5232 Villigen, Switzerland}
\affiliation{Elettra Sincrotrone Trieste, s.s. 14 km 163.5 in Area Science Park, 34149 Trieste, Italy}
\author{Nicholas~C.~Plumb}
\affiliation{Photon Science Division, Paul Scherrer Institut, CH-5232 Villigen, Switzerland}
\author{J.~Hugo~Dil}
\email[]{hugo.dil@epfl.ch}
\affiliation{Institut de Physique, \'{E}cole Polytechnique F\'{e}d\'{e}rale de Lausanne, CH-1015 Lausanne, Switzerland}
\affiliation{Photon Science Division, Paul Scherrer Institut, CH-5232 Villigen, Switzerland}
\author{Milan~Radovi\'{c}}
\email[]{milan.radovic@psi.ch}
\affiliation{Photon Science Division, Paul Scherrer Institut, CH-5232 Villigen, Switzerland}

\date{\today}

\begin{abstract}

The two-dimensional electron gas found at the surface of SrTiO$_3$ and related interfaces has attracted significant attention as a promising basis for oxide electronics. In order to utilize its full potential, the response of this 2DEG to structural changes and surface modification must be understood in detail. Here, we present a study of the detailed electronic structure evolution of the 2DEG as a function of sample temperature and surface step density. By comparing our experimental results with \textit{ab initio} calculations, we found that a SrO-rich surface layer is a prerequisite for electronic confinement. We also show that local structure relaxations cause a metal-insulator transition of the system around 135~K. Our study presents a new and simple way of tuning the 2DEG via surface vicinality and identifies how the operation of prospective devices will respond to changes in temperature.

\end{abstract}


\maketitle

\section{Introduction}

SrTiO$_3$ (STO) features amid the most popular transition metal oxides, being widely used as a substrate, buffer layer, and high dielectric medium due to its structural and electronic properties. The interest in STO and STO-based heterostructures boosted in the last years due to the plethora of intriguing properties found in these systems. Prominent examples are the 2-dimensional electron gas (2DEG) \cite{Ohtomo:2004}, and the giant spin-charge conversion found in the LaAlO$_3$/STO (LAO/STO) interface, as well as in STO \cite{Noel:2020}. These features render STO a crucial material for the implementation of oxide-based electronics \cite{Lorenz:2016}. Quickly after the report on the remarkable transport properties emerging at the LAO/STO interface \cite{Ohtomo:2004} and other STO-based heterostructures \cite{Moetakef:2012,Niu:2013,Raghavan:2015,vonSoosten:2019}, a 2DEG was directly observed by angle-resolved photoemission (ARPES) on the surface of bare STO \cite{Santander:2011,Meevasana:2011}, renewing the interest in the basic physics of this material.

STO is a wide-gap insulator that crystallizes in a cubic structure at room temperature in its bulk form \cite{Frederikse:1964,Lee:1975}. At T$_c$~=~105~K, STO goes through a second-order phase transition from cubic to tetragonal symmetry, doubling its unit cell \cite{Farrel:1964,Shirane:1969}. Although the softening of the phonon mode at the $\Gamma$-point during cooling down favors a polar ground state, quantum fluctuations prevent the system from actually becoming ferroelectric at low temperatures \cite{Kiat:1996,Riste:1971,Muller:1979,Zhong:1996}. With this behavior, STO remains on the verge of its paraelectric phase as incipient or weak ferroelectric \cite{Noguera:2000}. 

Intriguingly, it has been shown that the cubic-tetragonal phase transition of the surface layers occurs at a higher temperature (T$_c\sim$150~K) \cite{Salman:2006,Smadella:2009,Salman:2011}, indicating a surface atomic structure distinct from the bulk. 
Indeed, many different surface structures of STO has been observed by low-energy electron diffraction (LEED) \cite{vdHeide:2001,Bickel:1989}, reflection high-energy electron diffraction (RHEED) \cite{Hikita:1993}, medium energy ion scattering (MEIS) \cite{Ikeda:1999}, surface x-ray diffraction (SXRD) \cite{Charlton:2000,Herger:2007}, and grazing incidence x-ray scattering (GIXS) \cite{Salluzzo:2013}, but there is only little agreement among the extensive list of results. Although the techniques employed have different sensitivities and probing depths, all results show altered interlayer distances (referred to as surface relaxation) and off-centering of atoms in the same layer (known as rumpling). These distortions extend through a few unit cells \cite{Salluzzo:2013}.
The diverging experimental results make it difficult for a direct comparison with the calculated structures, which in turn show more consistent figures for the surface relaxation and rumpling. Although discrepancies are still present when comparing different methods and exchange-correlation functionals \cite{Heifets:2001}, most of the results seem to indicate that the formation of surface dipole moments is stable at the STO surfaces, notably in SrO-terminated slabs and at low temperatures \cite{Ravikumar:1995,Cheng:1997, Li:1998, Padilla:1998}. 

Concerning the metallic states on the STO surface, there is a particular focus on the role of oxygen vacancies in the formation of both 2DEG and the accompanying in-gap states (IGS) \cite{Santander:2011,Meevasana:2011,Walker:2015}, including tight-binding models, density functional theory (DFT) and dynamical mean-field theory (DMFT) calculations \cite{Ravikumar:1995,Eglitis:2008,Lechermann:2016,Gosh:2016}. Further ARPES  \cite{Chen:2015,Moser:2013,Wang:2015} and \textit{ab initio} \cite{Janotti:2014,Hao:2015} calculations show the strong interaction between the lattice and the electronic system, causing both large and small polarons, which are temperature and carrier density dependent.
More recently, resonant inelastic X-ray scattering (RIXS) data on STO and the LAO/STO interface confirmed the formation of large polarons in these materials \cite{Geondzhian:2020}. All these findings emphasize that the surface structure of STO and the underlying layers are strongly coupled to the 2D electronic system, and may even be dependent on each other.

ARPES is the most direct tool to study the 2DEG on the STO surface and can also give information about the surface order. The vast majority of ARPES studies of the 2DEG on STO were performed at fixed, low temperatures (typically 20~K), where the sample is nominally in the tetragonal phase. In this work, we investigate the 2DEG on Nb-doped STO wafers with temperature-dependent ARPES across the bulk and surface structural phase transitions and interpret the changes in the spectra with the aid of \textit{ab initio} calculations. We tracked the electronic structure's evolution through an extensive temperature range to identify the influence of the bulk phase transition, surface relaxation, and defects. Here we explicitly considered the role of the surface structure on the formation and properties of the 2DEG on STO. Our results show that the appearance of a 2DEG on STO (001) surface requires a SrO termination. More importantly, within a simplified slab model, we relate the 2DEG observed for surfaces with different step densities and at different temperatures to varying levels of structural distortion. Our study reveals that the 2DEG is very sensitive to surface structure distortions, and strongly influenced by temperature and surface vicinality.

\section{Results}
\subsection{Flat SrTiO$_3$ surface}

\begin{figure}
	\includegraphics[width=0.9\columnwidth]{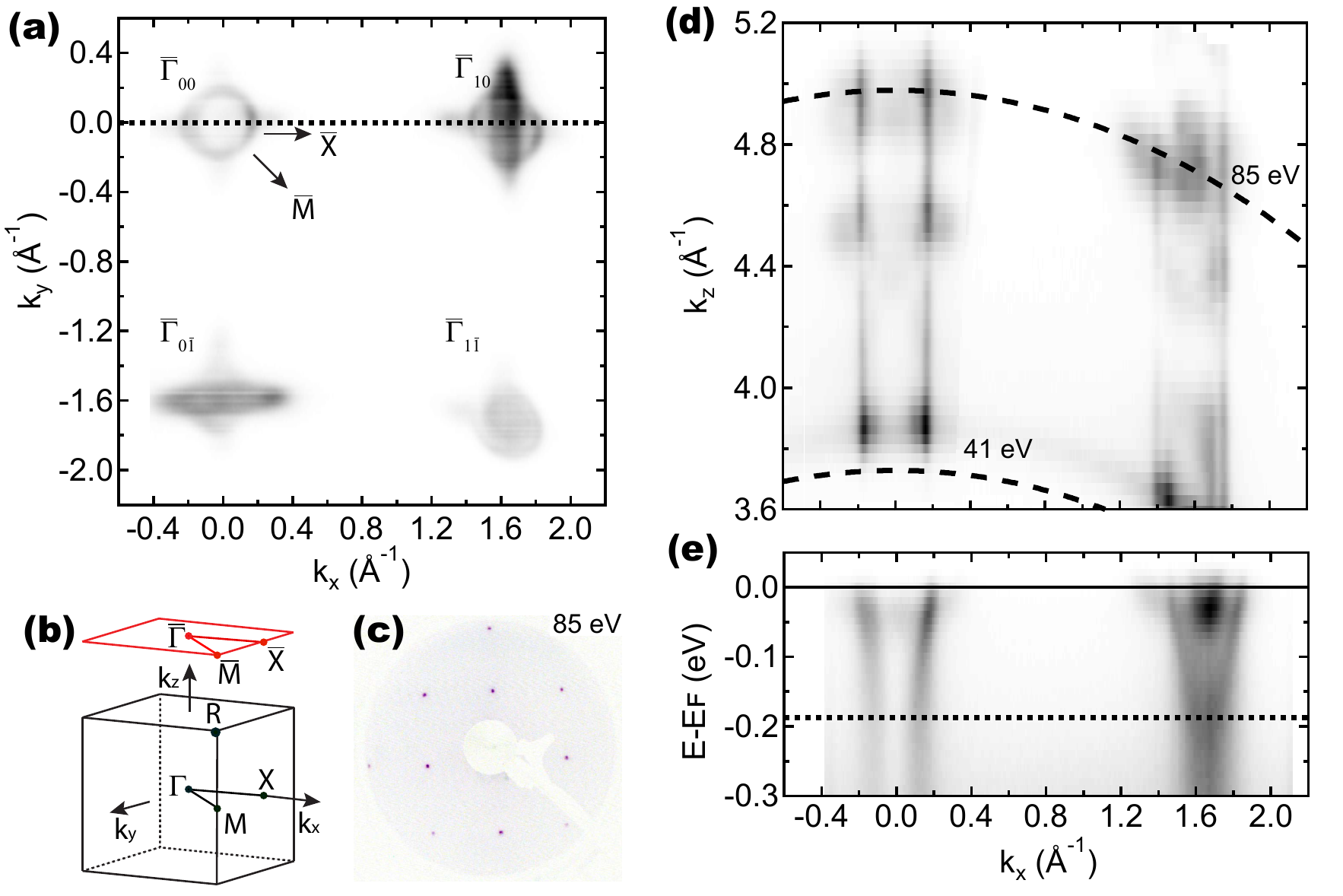} 
	\caption{\textbf{3D Fermi surface mapping of metallic state at SrTiO$_3$ (001) surface.} (a) Fermi surfaces map covering four Brillouin zones in the k$_x$-k$_y$ plane, measured with h$\nu$~=~85~eV and right-handed circular polarization (C$^+$). The arrows indicate the high-symmetry direction in (001) surface projection of the simple cubic Brillouin zone shown in (b). (c) LEED pattern showing a 1$\times$1 surface. (d) Fermi surfaces map covering $\overline{\Gamma_{00}}$- and $\overline{\Gamma_{01}}$-point in the k$_x$-k$_z$ plane, measured with C$^+$-polarized light. In the conversion to k$_z$, we used an inner potential V$_0$~=~14.5~eV, and the dashed lines indicate h$\nu$~=~41~eV and h$\nu$~=~85~eV. (e) Band dispersion maps obtained along the dashed line in (a).}
	\label{FS}
\end{figure}

At the first exposure of the sample to photons of 85~eV, we do not observe any intensity at the Fermi level (E$_F$). During the experiment parabolic states develop with a continuously increasing intensity, until the ARPES signal reaches saturation. All data shown were measured in this condition.

Figure~\ref{FS}(a) shows an extended Fermi surface map of a flat ($\leq$ 0.2\% miscut) STO(001) wafer, measured with circular polarized (C$^+$), h$\nu$=85~eV photons. The data spans four surface Brillouin Zones (SBZ), shown in red in Figure~\ref{FS}(b) and labeled accordingly. The distance between the points match with the in-plane reciprocal-lattice vector ($\lvert \textbf{G} \rvert \sim $ 1.6~\AA$^{-1}$), indicating the absence of any long-range electronic reconstruction, in agreement with the LEED pattern in Figure~\ref{FS}(c). It is worth noting, however, that a recent work employing non-contact atomic force microscopy pointed out that surface reconstructions on STO  sometimes cannot be detected by ARPES. \cite{Sokolovic:2020b} 

Figure~\ref{FS}(d) show the Fermi surfaces measured in the k$_z$-k$_x$ plane (for k$_y$=0, marked by the dashed line in Figure~\ref{FS}(a)), around $\overline{\Gamma_{00}}$ (at k$_x$=0) and $\overline{\Gamma_{10}}$ (at k$_x\sim$1.6~\AA$^{-1}$), obtained with circularly polarized (C$^+$) light. For the conversion from $h\nu$ to k$_z$ an inner potential $V_0$=14.5~eV was used~\cite{Plumb:2014}. The two pairs of straight lines correspond to the non-dispersive, 2D light bands with d$_{xy}$ character \cite{Plumb:2014}, split by a Rashba-like spin-orbit interaction \cite{Santander:2014}. The ellipsoidal-shaped features observed in the vicinity of the $\Gamma$-point (k$_z\sim4.8$~\AA$^{-1}$) correspond to the heavy bands with d$_{xz,yz}$ character \cite{Plumb:2014}.

Although metallic states are already formed, the intensity in the ARPES spectrum at the Fermi level is negligible for $h\nu\leq41$~eV. This observation indicates that the transition probability to the photoemission final states is very low, and these states can hardly be detected. Besides, this effect indicates that the Ti~3$p$-Ti~3$d$ resonant process (the onset of the Ti$^{4+}$~M edges is around 38~eV) is a crucial step for the observation of the 2DEG with photoelectron spectroscopy, as also suggested in Ref.~\cite{Walker:2015}. 

The drastic variation in intensity distribution around the different $\Gamma$-points [Figure~\ref{FS}(a)] and its dependence on photon energy is due to strong matrix elements effects \cite{Damascelli:2003}. Such effects also hinder the observation of the bottom of the d$_{xy}$ band in the first Brillouin zone, whereas it is observed at 190~meV in the adjacent zone, as seen by the electron dispersion in Figure~\ref{FS}(e). The estimated 2D carrier density of this band is $n_{2D}=5.56 \times 10^{13}$~cm$^{-2}$, with the electronic dispersion matching a parabolic band with $m^*=0.70 m_e$ (with $m_e$ the free electron mass). In turn, the ellipsoidal $3d_{xz}$- and $3d_{yz}$-derived bands match well with degenerate parabolas with E$_{xz,yz}$=40~meV and effective masses $m^*=10m_e$ and $m^*=0.24m_e$, respectively. 

\begin{figure}
	\includegraphics[width=1\columnwidth]{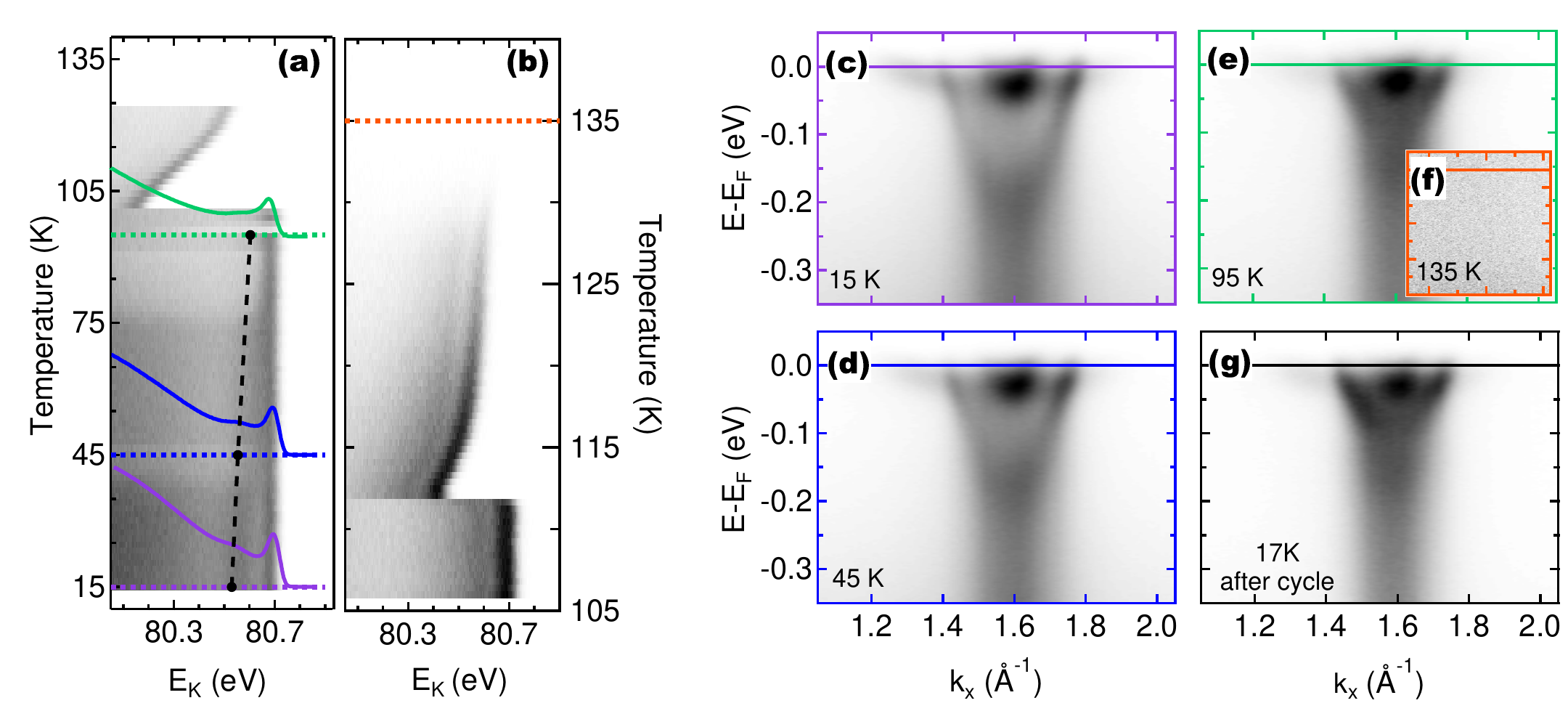} 
	\caption{\textbf{Temperature dependence of the 2DEG on STO (001) surface.} Temperature versus E$_K$ diagram showing the evolution of the 2DEG across temperature ramps 1 (a) and 2 (b), measured around $\overline{\Gamma_{10}}$ with $h\nu$=85~eV, C$^+$-photons. The dashed lines indicate the temperatures at which the  high-statistics band maps shown in (c-f) were acquired: 15, 45. 95, 135~ K, and then after the cycle at 17~K, respectively. The dashed lines in (c-f) indicate the bottom of the d$_{xy}$-derived band. Panel (a) also shows the energy distribution curves of (c-e), while each respective band bottom is represented with a circle.}
	\label{Tdep}
\end{figure}

For the temperature-dependent ARPES, shown in Figure~\ref{Tdep}, we focus on $\overline{\Gamma}_{10}$, using C$^+$ photons with h$\nu$=85~eV. Slow ($\sim$0.1~K/min) heating ramps were performed in order to keep the pressure in the experimental chamber well below 5$\times10^{-9}$~mbar, which was reached only at the highest temperature studied. Although the temperature ramps were performed very slowly, the error in the temperature reading is expected to be between 5 and 10\% due to thermal lag between the sample and the diode. Band dispersion maps were acquired continuously during the ramps, allowing us to track the position of the bands and of the Fermi level E$_F$ with temperature. Ramp 1 [Figure~\ref{Tdep}~(a)] was measured from 14~K to 125~K. Up to 100~K the position of E$_F$ remains at around 80.7 eV (in kinetic energy scale $E_K$). Close to 100~K it drastically shifts about 0.5~eV to lower E$_K$, then slowly moves towards the original position. After reaching 125~K, the sample was cooled down to 100~K and heated again to 145~K [ramp 2, Figure~\ref{Tdep}~(b)]. This time the ``jump'' of E$_F$ happened at around 112~K, a slightly higher value than observed during ramp 1. This change is attributed to a temperature offset acquired during the temperature cycle. We cannot exclude, however, the possibility that this is partially due to an intrinsic property of the sample. After the jump, E$_F$ smoothly moves back towards higher $E_K$, stabilizing around 20 meV below the initial value, until the metallic states vanish at around 135~K. 

In a photoemission experiment, the kinetic energy of E$_F$ is determined by the work function of the (grounded) analyzer, and shifts can only occur if there is an additional potential difference between the sample and the ground. Hence, the jump in $E_F$ indicates that at around 100~K the sample surface was suddenly subject to a more positive potential -- \textit{i.e.}, the sample tends to charge due to photoemission. This indicates a sudden increase in electrical resistance between the surface electron gas and the instrument ground, which points to an electrical phase transition, most likely in the underlying bulk or sub-surface region, that is linked with the tetragonal-cubic transition. At higher temperatures a discharge process helps the system to reach electrostatic equilibrium. The temperature of the jump is close to the bulk tetragonal-cubic phase transition in STO (105~K), and can be explained by the measured electrical transport behavior of Nb:STO crystals, which show a metal-insulator-metal transition in the 85-110~K range~\cite{SOM}. In this regard, the sudden change of the Fermi level's kinetic energy in Figures \ref{Tdep}~(a) and (b) can be understood as the response of the surface metallic states  to the bulk tetragonal-cubic phase transition.

Band dispersion maps were acquired continuously, while at selected temperatures, represented in Figure~\ref{Tdep} by dashed lines, the ramp was paused, and high-statistics maps were measured (the time evolution of the sample temperature is shown in Ref.~\cite{SOM}). 
For clarity, apart from the data in Figs.~\ref{Tdep}(a) and (b), E$_F$ has been set to zero binding energy in all other spectra. With increasing temperature, the d$_{xy}$ state progressively shifts to lower binding energies; at 15~K it is located at $\sim$~190~meV [Figure \ref{Tdep}(c)], with a small shift to $\sim$~166~meV at 45~K  [Figure \ref{Tdep}(d)], and a more pronounced change to 116~meV at T~=~95~K [Figure \ref{Tdep}(e)], until the metallic states vanish at 135~K [Figure \ref{Tdep}(f)]. These energy positions are plotted in [Figure \ref{Tdep}(a)], along with each respective momentum-integrated spectrum, where the upwards shift of the bottom of the d$_{xy}$ band can also be visualized. In turn, we could not observe a shift of the bottom of the heavy bands (50~meV) within the energy resolution of 6~meV. Consequently, the splitting $\Delta_{t2g}$ between the $d_{xy}$ and d$_{xz,yz}$ bands decreases from around 140~meV at 15~K to 66 meV at 95~K. The carrier density of the outer d$_{xy}$ band is monotonically reduced from $\sim$5.56 $\times 10^{13}$ cm$^{-2}$ at 15~K, to $\sim$3.58 $\times 10^{13}$ cm$^{-2}$ at 95~K, while the effective masses were found to fall in the 0.67 - 0.75 m$_e$ range.

\begin{figure}
	\includegraphics[width=0.8\columnwidth]{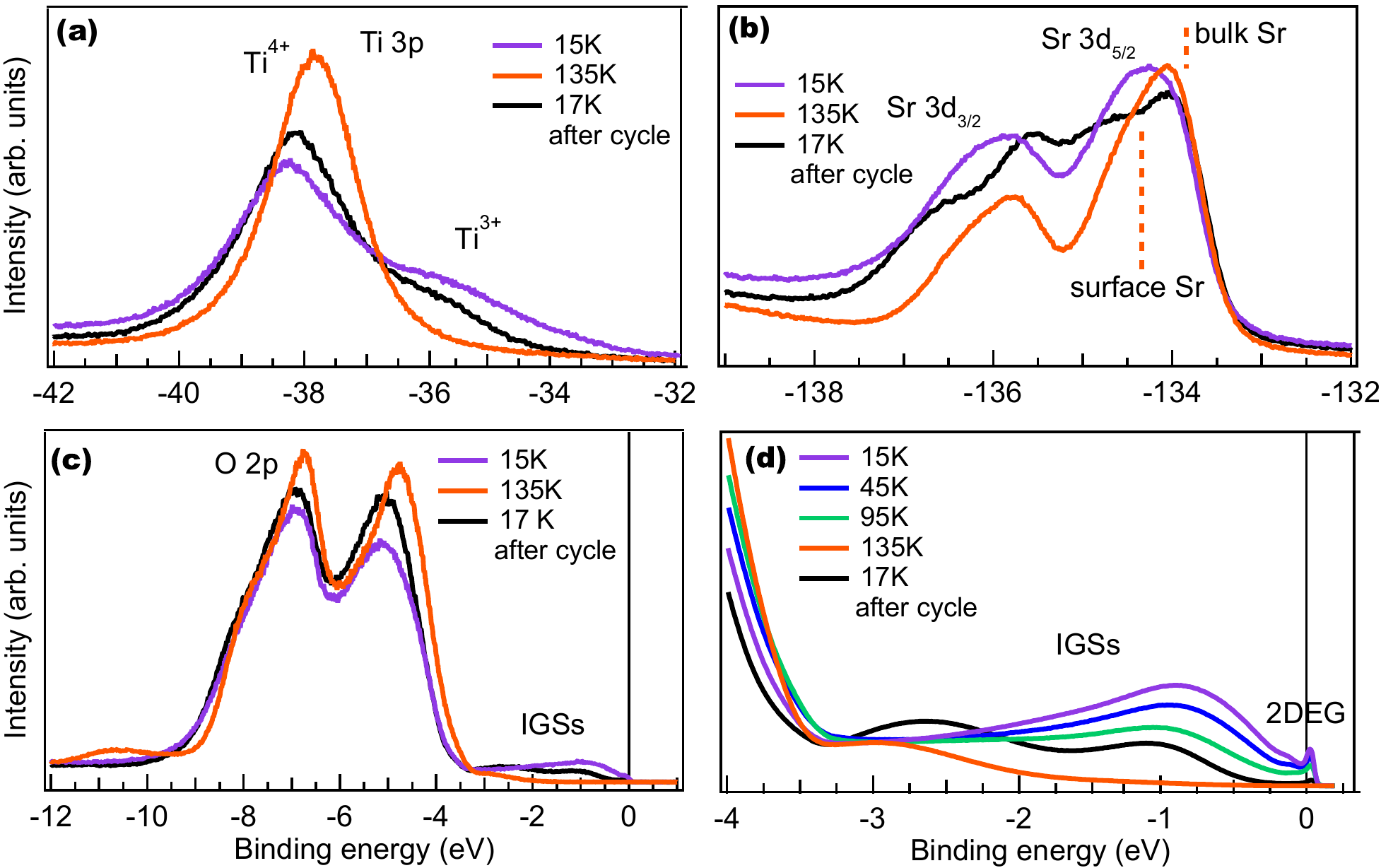} 
	\caption{{\textbf{Temperature dependence of the XPS of STO (001) wafer.} (a) Ti~3$p$ core-level measured at normal emission at low (15~K) and high (135~K) temperatures, as well as at 17~K after the cycle. (b) Equivalent data for the Sr~3$d$ core-level and (c) the valence band. (d) Evolution of the IGSs measured at 15, 45, 95 and 135~K, as well as 17~K.}}
	\label{XPS}
\end{figure}

Figure \ref{XPS} shows core-level and valence band X-ray photoemission spectroscopy (XPS) using $h\nu$ = 170~eV  measured at 15 and 135~K, \textit{i.e.} the base temperature and the one at which the 2DEG has just vanished. At low temperature (purple), the Ti~3$p$ core-level features the main Ti$^{4+}$ peak and the Ti$^{3+}$ shoulder, associated with the formation of in-gap and metallic states \cite{Plumb:2014}, while the Sr 3$d$ spectrum shows the characteristic surface and bulk contributions (dashed lines) reported for thin films~\cite{Guedes:2020}. The calculated Sr/Ti ratio, taking into account each photoionization cross-section \cite{Yeh:2000}, is 1.41, which falls in the SrO-terminated range reported in Ref.~\cite{Rebec:2019}. At 135~K (orange), the Ti$^{3+}$ shoulder vanishes and the Ti$^{4+}$ peak shifts $\sim$480 meV to lower binding energy, while the Sr 3$d$ peaks shift $\sim$340 meV to lower binding energy. 

Changes in shape are also seen across the valence band [Fig.\ref{XPS}(c)], mainly composed of O~2$p$ states from 3 to 8~eV below the Fermi level, with an additional peak around -11 eV at 135~K, usually assigned to OH$^-$ adsorption \cite{Yukawa:2013}. Additionally, the peaks in the O~2$p$ band become sharper, and a shift of $\sim$340~meV towards E$_F$ is observed. Closer to $E_F$, the in-gap states (IGSs) [Fig~\ref{XPS}(d)] are likely composed of several overlapping peaks, with local maxima around 1.3 and 2.8~eV which we refer to as low- and high-binding energy states. Despite the broad features, we observe that with increasing temperature the low-binding energy state shifts away from E$_F$ by approximately 180 meV and becomes gradually weaker until its suppression at 135~K. The high-binding energy state also shifts further down to higher binding energy (about 800 meV) while only showing a small reduction of intensity, which we assign to the reduced tail of the low-binding energy IGS. 

After completing ramp 2, the sample was once again cooled down to low temperature (17~K). The electronic band dispersion shown in Figure~\ref{Tdep}(g) reveals that a band is filled only around 100 meV, with a carrier density of 3.35$\times 10^{13}$cm$^{-2}$ (similar to that observed around 95~K in ramp 1). The Ti 3$p$, Sr 3$d$, and valence band XPS spectra show differences with respect to the first measurements at 15~K. While the total Sr to Ti ratio remains almost unchanged at 1.37, the intensity of the Ti$^{3+}$ is reduced in the Ti 3$p$ spectrum, and the Sr 3$d$ core-level broadens and changes its shape, seemingly gaining an additional component. This indicates the presence of an additional chemically different Sr atom, but whether this is related to the presence of OH$^-$ combined with UV irradiation at 135~K would require further investigations that go beyond the scope of this manuscript.  In the valence band, the low- and high-binding energy IGSs shift further to lower binding energy, with a  pronounced increase in the intensity of the second. 

The IGSs have been associated with the existence of different kinds of point defects in the crystal, including oxygen vacancies, interstitial oxygen, Ti--Nb anti-sites, and Sr-O vacancy complexes \cite{Choi:2009,Kim:2009,Chambers:2018,AlZubi:2019}. Such defects can electrostatically trap charges such as in doped semiconductors or, more intricately, reflect the formation of small polarons, quasiparticles arising due to strong, short-range electron-phonon interaction, which show a typical binding energy of 1~eV \cite{Fujimori:1996,Jeschke:2015,Janotti:2014,Hao:2015}. The low- and high-binding energy IGSs in Fig~\ref{XPS}(d) exhibit significantly different temperature evolution, which indicates they do not have the same physical origin.
More importantly, the complete disappearance of the low-binding energy IGS at higher temperatures can only be explained by a weakening of the electronic trapping mechanism around the defect, rather than by the disappearance of the defect itself.

It is also intriguing that the low-binding energy IGS and metallic states vanish at the same temperature, which can be related to the reported surface structural phase transition at 150~K \cite{Salman:2006,Smadella:2009,Salman:2011}. This may indicate that some particular surface structure (somehow related to the tetragonal phase) is crucial both for stabilizing the metallic states and the small polarons. Moreover, the distinct chemical shifts observed in the Ti~3$p$, Sr~3$d$, and valence band indicate a site-specific component of the total energy shift, which suggests a structural origin.

\subsection{Stepped SrTiO$_3$ surfaces}

\begin{figure}
	\includegraphics[width=1\columnwidth]{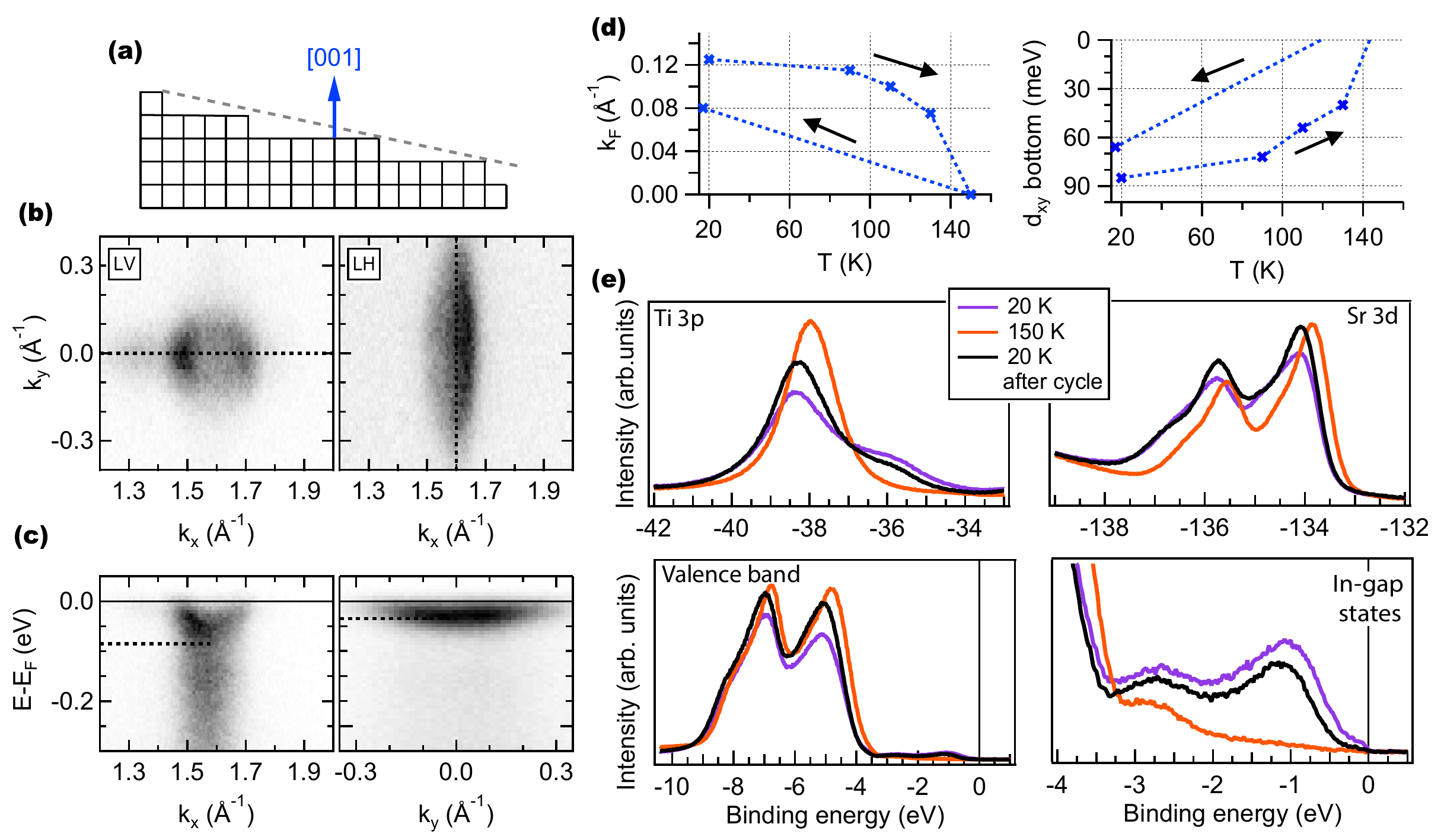}
	\caption{\textbf{2DEG on 10\degree-miscut STO sample.} (a) Illustration of the 10\degree\ miscut to the (001) surface with resulting terrace sizes. (b) Fermi surface measured around  $\overline{\Gamma_{10}}$ with $h\nu$=85~eV, LV- and LH-polarized light and (c) corresponding band dispersion maps acquired along the dashed lines in (b). (d) k$_F$ and band bottom of the d$_{xy}$ state for different temperatures. (e) Temperature-dependent XPS spectra of Ti~3$p$ and Sr~3$d$ core-levels, valence band and IGSs.}
	\label{vicinal}
\end{figure}

To further understand the observed changes in 2DEG during the temperature cycle, we performed the same set of photoemission experiments on a stepped STO wafer with a 10\degree\ miscut in respect to the [001] direction. This sample is expected to show atomic terraces of 22~\AA~[Figure \ref{vicinal}(a)], in contrast to the typical 500 - 2000~\AA~ found on flat STO(001) crystals. This high density of step edges grants the surface an additional degree of freedom for relaxation. As seen in the Supplemental material~\cite{SOM}, both this sample and ones with a 5\degree\ miscut also host a 2DEG without any signs of surface reconstruction, although the data for these samples are less sharp than for the flat one due to the increased incoherent scattering of photoelectrons off the step edges.

Matrix element effects allow us to probe states with different orbital characters, depending on the chosen light polarization. In our experimental geometry, using linear vertical (LV) polarization probes states with d$_{xy}$ and d$_{yz}$ character, while linear horizontal (LH) polarization highlights the d$_{xz}$-derived bands \cite{Plumb:2014}. The Fermi surfaces measured with h$\nu$~=~85 eV and each light polarization are shown in Figure~\ref{vicinal}(b), while Figure~\ref{vicinal}(c) show the band dispersion maps along the dashed lines in Figure~\ref{vicinal}(b). The bottom of the d$_{xy}$ band is located around 85 meV, while the bottom of the d$_{xz}$ band is around 35~meV. A small shift associated with different electron affinity on the stepped surface is expected, which may account for the 15~meV shift of the heavy bands relative to the flat surface. However, the large upward shift of around 110 meV of the d$_{xy}$ band suggests that the extra degree of freedom provided by the stepped surface strongly affects the band filling of the 2DEG, making the step density a viable knob to tune it.

The temperature dependence of the 2DEG on this sample is summarized in Figure~\ref{vicinal}(d). Similarly to flat STO, the d$_{xz}$ state is found to not shift as a function of temperatures, whereas the d$_{xy}$ band shifts towards E$_F$. The variation of the splitting $\Delta_{t2g}$ and k$_F$ of the d$_{xy}$ band with temperature is monotonic, until the disappearance of the metallic states at around 150~K. The sample was then cooled down to 20~K, after which we again observe a shallower band bottom and smaller k$_F$ than previously observed at the same temperature. Finally, the Sr~3$d$, Ti~3$p$, and valence band XPS spectra measured at 20, 150, and 20~K after the temperature cycle, are shown in \ref{vicinal}(e). The data reveals a behavior very similar to the flat STO, with the suppression of the Ti$^{3+}$ component of the Ti 3$p$ core level with temperature, a change in the shape of the Sr 3$d$ core-level, and the disappearance of the low-binding energy IGS concomitantly with the Ti$^{3+}$ and metallic states.

Along with the reduced occupied bandwidth found in the 10\degree-miscut STO, the similar evolution with temperature for both flat and stepped samples, namely the irreversible change in occupied bandwidth of the d$_{xy}$ band, as well as the non-rigid core-level energy shifts and intensity ratio changes after the temperature cycle, suggest that the 2DEG is intimately connected to the particular surface structure.

\subsection{\textit{Ab initio} calculations}

Theoretically, structural effects on the electronic band structure can be captured by density functional theory (DFT) calculations, which we employed to evaluate whether surface structural changes could be responsible for the observed changes in the 2DEG band structure. Our strategy was to compare the measured $\Delta_{t2g}$ at different temperatures with the ones obtained from DFT calculations for STO slabs with different crystal structures as described below.

We used slab models with both TiO$_2$ and SrO terminations, but only the SrO-terminated slab resulted in a band structure where the lowest-lying surface-derived d$_{xy}$ band is below the bulk-derived counterpart~\cite{SOM}. This means that the formation of the 2DEG is favored on the SrO termination, as already shown theoretically~\cite{Delugas:2015} and experimentally~\cite{Rebec:2019, Guedes:2020}, and in agreement with our XPS results. Henceforth we refer only to SrO-terminated slabs. To focus on the role of the surface atomic structure on the band structure, we opted to not include oxygen vacancies in our calculations.

\begin{figure}
	\includegraphics[width=0.8\columnwidth]{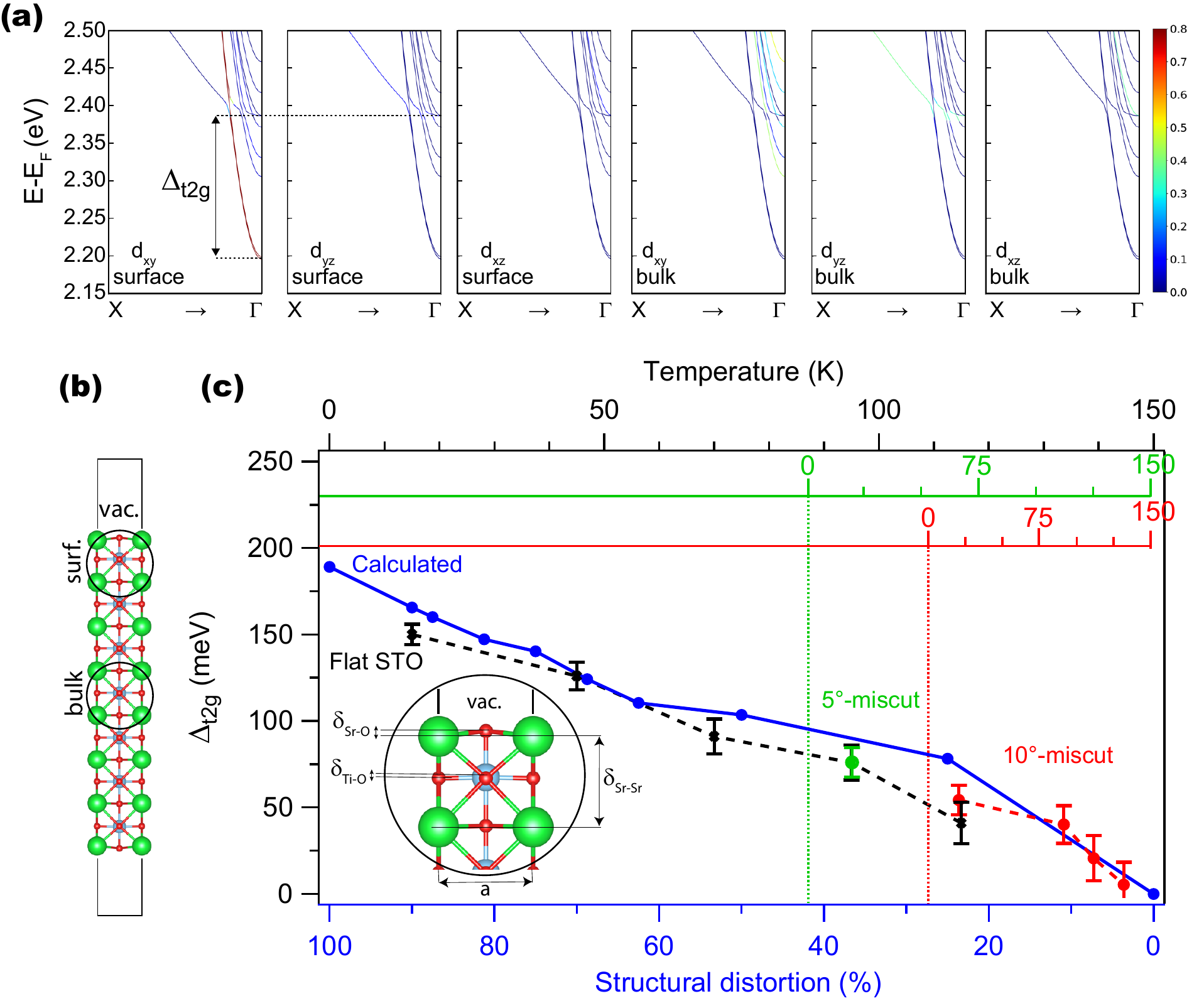}
	\caption{\textbf{\textit{Ab initio} calculations.} (a) Electronic band structures of relaxed 1$\times$1$\times$5, SrO-terminated STO slab, shown on (b), highlighting different orbital projections (xy, xz, yz) and origins (bulk and surface). (c) Comparison between calculated $\Delta_{t2g}$ for different degrees of distortion (blue) and measured $\Delta_{t2g}$ for flat, 5\degree- (green) 10\degree-miscut (red) STO at different temperatures. The origins of the green and red temperature axes were shifted to match the closest point for the flat sample, see text for details. The inset further details the distortions present at the surface layer. The crystal structure was generated with the VESTA software \cite{Momma:2011}. Sr atoms are represented by green spheres, O by red and Ti by light blue spheres.}
	\label{dft}
\end{figure}

Figure~\ref{dft}(a) shows the calculated band structure along the X$\Gamma$ direction of our relaxed 1$\times$1$\times$7 STO slab. The structure is depicted in Figure~\ref{dft}(b), highlighting the bulk and surface layers. The band structure reveals an electron-like band with Ti-d$_{xy}$ character almost entirely originating from the surface layer, as well as the d$_{xz}$ and d$_{yz}$ bands with a mixed bulk-surface origin, in good agreement with previous works~\cite{Plumb:2014}. The calculated $\Delta_{t2g}$ amounts to 189 meV, approaching our observed splitting of 150 meV, and in good agreement with values reported in previous studies~\cite{Santander:2011,Meevasana:2011,Plumb:2014}. However, the upper branch of the d$_{xy}$ band, arguably the Rashba-pair of the lower branch \cite{Santander:2014}, was not captured by the calculations since we do not include the spin-orbit interaction. We note that the surface-derived d$_{xy}$ band shows a small breaking of degeneracy of 3~meV, which we ascribe to finite-size effects and we further address this point in \cite{SOM}.

From the relaxed SrO-terminated slab, we computed the atomic displacements concerning each respective bulk-truncated (unrelaxed) structure and used them to generate slabs with different degrees of distortion with fractions of the final displacements, whose total energies and band structures were calculated. The inset of Figure~\ref{dft}(c) shows a zoom around the surface layer, where the atomic displacements $\delta_{Sr-Sr}$, $\delta_{Sr-O}$ and $\delta_{Ti-O}$ are defined, and $a$ is the relaxed in-plane lattice parameter. Changing the degree of distortion will alter these values, which are described in~\cite{SOM}. In DFT, all calculations are performed at 0~K, for which the relaxed structure provides the minimum energy. However, the model structures in our scheme correspond to higher energy configurations accessible at higher temperatures. This model aims to show how one effective parameter, a set of atomic displacements following a parabolic path in the potential energy surface at 0~K \cite{SOM}, affects $\Delta_{t2g}$. Although we capture the correct trend within this simple model without defects, electron-phonon interaction, and light, we are not able get the correct energy scale involved in the formation and temperature evolution of the 2DEG. Therefore, we decided to use direct comparison with the experimentally observed $\Delta_{t2g}$ to assign each corresponding temperature to a different structural distortion.

Figure~\ref{dft}(c) shows the calculated $\Delta_{t2g}$ (blue markers) for selected degrees of distortion (bottom horizontal axis), where we see that the splitting monotonically decreases with distortion. For direct comparison, we plot the experimentally observed splittings as a function of temperature (top horizontal axis) for the flat STO sample of Figure~\ref{Tdep} (black markers), which nicely correspond to the behavior and magnitude of the calculated $\Delta_{t2g}$. In addition, we plot the observed splittings for the 10\degree-miscut STO sample of Figure~\ref{vicinal} (red markers), as well as the splitting measured at 20~K for a 5\degree-miscut STO sample (green marker) \cite{SOM}, with a shift in the origin of their temperature axis to match he corresponding $\Delta_{t2g}$ of the flat STO sample. This analysis also relates the splitting found in stepped surfaces with a respective degree of distortion. The good match with the calculated trend indicates that the differences found in the 2DEG on flat and stepped STO surfaces can be attributed to different surface atomic structures, likely induced by the extra degree of freedom for structural relaxation in the miscut surfaces.

Due to the symmetry of the slab, atomic displacements are restricted to the z-direction only. These may either cause a change in the surface lattice parameter ($\delta_{Sr-Sr}$) with regard to the bulk value $a$, or induce interatomic displacements ($\delta_{Sr-O}$ and $\delta_{Ti-O}$). We have also investigated the band structure of SrO-terminated 1$\times$1$\times$6 slabs of STO in the AFD phase, although we do not observe the doubled unit cell characteristic of this crystallographic phase in our LEED and ARPES results. Upon decreasing the lattice distortion, which in this case corresponds to the AFD angle, the calculations reveal the same trend as observed in the cubic case. This indicates that the polar (out-of-plane) atomic displacements play a key role in generating the observed band structure. Further details on the AFD calculations can be found in Supplemental material~\cite{SOM}.

\section{Discussion and summary}

Our ARPES and DFT studies show that the SrO termination for STO (001) surface is a requirement for the creation of the 2DEG, which was suggested theoretically~\cite{Delugas:2015} and observed in the study with MBE-grown STO films of Rebec \textit{et al.}~\cite{Rebec:2019}. More precisely, our DFT calculations found that the relaxed structure with a SrO termination results in a band structure with  correct orbital character ($xy$, $xz$, or $yz$) and layer origin (bulk- or surface-derived), as well as a closely matching $\Delta_{t2g}$, when compared to the ARPES data. Further, the experimental variation of $\Delta_{t2g}$ with increasing temperature can be captured by atomic displacements, suggesting that surface structural distortions are responsible for determining the properties of the 2DEG on STO. We expect that further improvements in the theoretical treatment of the system will lead to a more accurate description of the crystal and band structures with regard to which type and degree of distortion corresponds to each value of $\Delta_{t2g}$. 

In the above discussion we did not consider any particular source of electron doping. The abundance of charges generated during the photoemission experiment grants the availability of free charges to occupy the empty Ti t$_{2g}$ states at the surface. In our model, the primary role of the oxygen vacancies, in this case, is not of a charge donor, but instead of a promoter of structural distortions, necessary for the stabilization of the polaronic states \cite{Janotti:2014,Hao:2015,Geondzhian:2020}. This effect can be understood in more detail by considering the fact that the presence of an electric field will induce a polar atomic distortion at the surface of STO \cite{Khalsa:2012}. Photoexcited electrons will be a natural source of this electric field, as are charges from other sources such as overlayers. This polar distortion becomes stabilized by trapping the electron and thus forming a large polaron, similar to the trapping and stabilization of small polarons like the in gap states. Once the trapped charge density becomes high enough, any field can be effectively screened and the process saturates \cite{Khalsa:2012}, explaining the quasi-universal band filling of the 2DEG \cite{Santander:2011}.

In this picture it is clear that DFT can't reproduce the band filling of the 2DEG, because it is not the ground state of the system, but rather a meta stable state. On the other hand, the obtained $\Delta_{t2g}$ in a first approximation only depends on the local structure and can thus be captured by DFT. That the structural distortion depends on the sample temperature can be understood from the dependency of the polar structural distortion and polaron stabilization on the electron phonon coupling. The temperature will have a direct impact on the coupling constant, and an indirect impact (through a different equilibrium structure) on the phonon modes' frequencies and amplitudes. Similarly, the presence of step edges significantly alters the available phonon modes, and typically leads to a softening \cite{Nie:1995,Witte:1995}. This also explains why a larger vicinality is equivalent to an increase in temperature on the flat surface.

To conclude, we have shown that the changes of the 2DEG found on STO surfaces as a function of sample temperature and surface step density can be explained by considering the structural relaxation of the surface layer, strongly influenced by temperature and step edges. Our results show that the step density is a viable way of tuning the 2DEG on STO. Furthermore, our results provide additional evidence that the SrO surface termination has to be considered to explain the presence of the 2DEG, and suggest a temperature range where prospective STO-based devices potentially operate. These findings will help to steer the engineering of orbital and lattice degrees of freedom in oxide-based electronics.

\section{Acknowledgments}
This work was supported by the Swiss National Science Foundation (SNF) Project No. PP00P2\_144742 and No. PP00P2\_170591. M.R. and E.B.G. acknowledge the support of SNF Project No. 200021\_182695. W.H.B. acknowledges the Pr\'{o}-Reitoria de Pesquisa of Universidade Federal de Minas Gerais, and the National Laboratory for Scientific Computing (LNCC/MCTI, Brazil) for providing HPC resources of the SDumont supercomputer, which have contributed to the research results, URL: http://sdumont.lncc.br.

\section{Methods\label{methods}}

\subsection{Sample preparation and angle-resolved photoemission}

For this study, commercially available 0.05 w.t.\% Nb-doped STO substrates (SurfaceNet GmbH) with miscuts of $\leq 0.2\degree$, 5\degree~ and 10\degree~to the nominal (001) surface were used. The samples have a size of 5$\times$10$\times$0.5 mm$^3$ and are cut and polished by the manufacturer. The samples are submerged in deionized water for approximately 30 min and then etched in buffered HF solution for 30 seconds. After this step, the samples are washed in a sequence of baths in deionized water to terminate the etching. Subsequently, the samples are dried by annealing them in a constant high-purity oxygen flow to 1000\degree~C. After the ex-situ treatment, the samples are annealed to 550\degree~C in 100 mbar of O$_2$ in-situ. Finally, the sample is low temperature annealed at 300\degree~C for approximately 12 hours. After this treatment, the samples were in situ transferred to the ARPES manipulator and measured without further treatment. All the data presented were measured at the high-resolution angle-resolved photoemission spectroscopy endstation at Surface and Interface Spectroscopy (SIS) beamline of the Swiss Light Source. The photoelectron analyzer in use is a Scienta R4000 hemispherical analyzer. The sample is cooled by a liquid helium cryostat which allows measurements at temperatures as low as 15~K, while higher temperatures were achieved by slowly heating the cryostat away from the sample. This method allows changing the temperature without significantly influencing the pressure, but also creates the thermal lag discussed in the main text.  

\subsection{\textit{Ab initio} calculations}

Our density functional theory calculations were performed within the Perdew-Burke-Ernzehof generalized gradient approximation (PBE-GGA)~\cite{PBE}, using projector augmented wave (PAW) potentials~\cite{paw}, as implemented in the Vienna \textit{Ab initio} Simulation Package (VASP)~\cite{vasp1,vasp2}. In addition, we employed the DFT+U functional of Liechtenstein {\it et al.}~\cite{lich_1998} with $U = 5$ eV and $J = 0.64$ eV, as similarly performed in Ref.~\cite{Shen:2012}. A basis set of 500 eV were used, and the structures were relaxed until the forces on atoms were less than 0.01 eV/\AA{}. The relaxation of atoms were done using a 4 $\times$ 4 $\times$ 1 \textit{k}-mesh, whereas the band structures were evaluated using a 8 $\times$ 8 $\times$ 2 \textit{k}-points set.

\bibliographystyle{apsrev4-1}

\bibliography{ReferencesFull}

\pagenumbering{gobble}

\end{document}